\documentclass[11pt,fleqn]{article}
\usepackage{amsmath}
\usepackage{bbm}

\usepackage{amsmath,amssymb,amsthm,enumerate,cite}

\newcommand{\const}{\mathop{\rm const}\nolimits}

\newcommand{\pr}{\mathop{\rm pr}\nolimits}
\newcommand{\thetbn}{\arabic{nomer}}

\newcommand{\ord}{\mathop{\rm ord}\nolimits}

\newcommand{\arctanh}{\mathop{\rm arctanh}\nolimits}

\newcounter{tbn}

\newtheorem{theorem}{Theorem}

\newtheorem{proposition}{Proposition}
{\theoremstyle{definition}

\begin{document}

\par\noindent {\LARGE\bf Reduction operators and exact solutions of variable coefficient nonlinear wave
equations with power nonlinearities  \par} {\vspace{4mm}\par\noindent
{\bf  Ding-jiang Huang~$^{\dag,\ddag,\S}$,Qin-min Yang~$^\dag$ and Shui-geng Zhou~$^{\ddag,\S}$}
\par\vspace{2mm}\par} {\vspace{2mm}\par\noindent {\it
$^\dag$~Department of Mathematics, East China University of Science and Technology,  Shanghai,\\
$\phantom{^\dag}$~200237, China\\
}} {\noindent \vspace{2mm}{\it $\phantom{^\dag}$~e-mail:
djhuang.math@gmail.com,~djhuang@fudan.edu.cn
}\par}

{\par\noindent\vspace{2mm} {\it $^\ddag$~School of Computer Science,
Fudan University,  Shanghai, $\phantom{^\dag{}^\ddag}$~200433,
China}}

{\par\noindent\vspace{2mm} {\it $^\S$~Shanghai Key Lab of
Intelligent Information Processing, Fudan University, Shanghai,\\
200433, China }}

{\vspace{5mm}\par\noindent\hspace*{8mm}\parbox{140mm}{{\bf Abstract}\\\small
Reduction operators, i.e. the operators of nonclassical (or conditional)
symmetry of a class of variable coefficient nonlinear wave equations with power nonlinearities is investigated within the
framework of singular reduction operator. A classification of regular reduction operators is performed
with respect to generalized extended equivalence groups. Exact solutions of some nonlinear wave model which are
invariant under certain reduction operators
are also constructed.
}\par\vspace{5mm}}

\section{Introduction}
In this paper, we study reduction operators, i.e., the operators of nonclassical (or conditional)
symmetry associated with a class of variable coefficient nonlinear wave equations with power nonlinearities of the form
\begin{equation}\label{eqVarCoefNonLWaveEq}
f(x)u_{tt}=(g(x)u^nu_x)_x+h(x)u^m,
\end{equation}
where $f=f(x), g=g(x)$ and $h=h(x)$ are three arbitrary functions, $fg\neq
0$, $n$ and $m$ are arbitrary constants, $t$ is the time coordinate
and $x$ is the one-space coordinate. The linear case is excluded
from consideration because it was well-investigated. We also assume the
variable wave speed coefficient $u^n$ to be nonlinear, i.e. $n \neq 0$. The case $n
=0$ is quite singular and will be investigated separately.

Many specific nonlinear wave models describing a wide variety of
phenomena in Mechanics and Engineering such as the flow of one-dimensional gas, shallow water
waves theory, longitudinal wave propagation on a moving
threadline, dynamics of a finite nonlinear string,
elastic-plastic materials and electromagnetic transmission line and so on, can be reduced to equation \eqref{eqVarCoefNonLWaveEq}
(see \cite{Ames1965/72} p.50-52 and \cite{Ames1981}).
Since 1970s, Lie symmetries and invariant solutions of
various kinds of quasi-linear wave equations in two independent variables that
intersect class \eqref{eqVarCoefNonLWaveEq} have been investigated \cite{Bluman&Cheviakov2007,Pucci&Salvatori1986,Torrisi&Valenti1985,Donato1987,
Ibragimov&Torrisi&Valenti1991,Ibragimov1994V1,Oron&Rosenau1986,Chikwendu1981,Arrigo1991,Gandarias&Torrisi&Valenti2004,Pucci1987,
Bluman&Temuerchaolu&Sahadevan2005,Bluman&Kumei1989,Huang&Ivanova2007,Huang&Mei&Zhang2009,Huang&Zhou2010,Huang&Zhou2011,Vasilenko&Yehorchenko2001} because of
the importance of the wave equation for various applications.
Recently, we have present a complete Lie symmetry and conservation law classification of class \eqref{eqVarCoefNonLWaveEq} \cite{Huang&Yang&Zhou2010a,Huang&Yang&Zhou2010b}. Classical Lie symmetry reduction and invariant solutions of some variable coefficients wave models which are singled out from the classification results are also investigated \cite{Huang&Yang&Zhou2010a}.

In general, using classical Lie symmetries reduction of partial differential equations can provide part of exact solutions of these equations\cite{Olver1986,Ovsiannikov1982}. In order to find more other types exact solutions,
one should generalize Lie's original reduction. The first approach to such generalization was present by Bluman and Cole in 1969 \cite{Bluman&Cole1969} (see also \cite{Bluman&Kumei1989}) in which they introduced a wider class of infinitesimal generators than
Lie symmetries. Later such infinitesimal generators were named nonclassical symmetries \cite{Levi&Winternitz1989} or conditional
symmetries \cite{Fushchych&Shtelen&Serov1993,Fushchych&Shtelen&Serov&Popovych1992,Zhdanov&Tsy&Pop1999,Fushchych&Zhdanov1992,Kun&Pop2009}, and were also extended by many authors to some concepts such as weak symmetry \cite{Olver&Rosenau1987} or differential constraints, etc., \cite{Olver1994,Olver&Rosenau1986,Pucci&Saccomandi1992,SSN1984,Yanenko1984}. Recently, Popovych et.al named it as `reduction operators' and present a novel framework, namely singular reduction
operators or singular reduction modules \cite{Kun&Pop2008,Boyko&Kunzinger&Popovych2012}, for finding an optimal way of obtaining the determining
equation of conditional symmetries. As application, they have investigated the
properties of singular reduction operators for a number of (1+1)-dimensional evolution equations and a specific wave equations \cite{Kun&Pop2008,Pocheketa&Popovych2011,Pocheketa&Popovych2012,Popovych2008,Popovych&Veneeva&Ivanova2007,Veneeva&Popovych&Sophocleous2009,Veneeva&Popovych&Sophocleous2011,Boyko&Popovych2013} by using this new framework.
However, for more general nonlinear wave equation \eqref{eqVarCoefNonLWaveEq}, there exist no general results.
In this paper, we employ Popovych's singular reduction operators theory to investigate the properties of nonclassical symmetries of class \eqref{eqVarCoefNonLWaveEq}.
We propose a complete classification of regular reduction operators for \eqref{eqVarCoefNonLWaveEq}
with respect to generalized extended equivalence groups and construct some non-Lie exact solutions for the nonlinear wave model which are
invariant under certain reduction operators. Below,
following \cite{Kun&Pop2008} we use the shorter and more natural term `reduction operators' instead of `operators
of conditional symmetry' or `operators of nonclassical symmetry'.

The rest of paper is organized as follows.
In Section \ref{SectiononNonSymmCla}, singular reduction operators, and in particular regular reduction operators
classification for the class under consideration are investigated. Section \ref{SectiononLieandNonSymmRedu} contains nonclassical
symmetry reduction of some nonlinear wave models. New non-Lie
exact solutions of the models are constructed by means the reduction.
Conclusions and discussion are given in section \ref{SectionOnConclusion}.

\section{Nonclassical symmetries}\label{SectiononNonSymmCla}

Nonclassical symmetries of class (\ref{eqVarCoefNonLWaveEq}) is
performed in the framework of the singular reduction operator \cite{Kun&Pop2008}.
All necessary objects (singular and regular reduction operator, etc.) can be found there \cite{Kun&Pop2008}. Before we proceed the investigation,
we can first simplify the class (\ref{eqVarCoefNonLWaveEq}).
Using the transformation
\begin{equation}\label{TranofEq}
\tilde t = t,\quad \tilde x =\int \frac{dx}{g(x)}, \quad \tilde u = u
\end{equation}
from theorem 1 in \cite{Huang&Yang&Zhou2010a}, we can
reduce equation \eqref{eqVarCoefNonLWaveEq} to $
\tilde f(\tilde x)\tilde u_{\tilde t\tilde t}=(\tilde u^n\tilde
u_{\tilde x})_{\tilde x}+\tilde h(\tilde x)\tilde u^m,
$
where $\tilde f(\tilde x) = g(x)f(x), \tilde g(\tilde x) = 1$ and
$\tilde h(\tilde x) = g(x)h(x)$. Thus, without loss of generality we can restrict ourselves to
investigation of the equation
\begin{equation}\label{g11eqVarCoefNonLWaveEq}
f(x)u_{tt}=(u^nu_x)_x+h(x)u^m.
\end{equation}
For convenient, we can further rewrite it as the form
\begin{equation}\label{g1eqVarCoefNonLWaveEq}
L[u]:=f(x)u_{tt}-(u^nu_x)_x-h(x)u^m=0.
\end{equation}
All results on symmetries and solutions of class \eqref{g11eqVarCoefNonLWaveEq} or \eqref{g1eqVarCoefNonLWaveEq} can be extended to class \eqref{eqVarCoefNonLWaveEq}
with transformations \eqref{TranofEq}.

According to the algorithm in
\cite{Kun&Pop2008}, we seek a
{\sl reduction operator} of class \eqref{g1eqVarCoefNonLWaveEq}  in the form
\begin{equation}\label{VectorFieldofNonClaSymm}
Q=\tau(t,x,u)\partial_t+\xi(t,x,u)\partial_x+\eta(t,x,u)\partial_u,~~~(\tau,\xi)\neq(0,0),
\end{equation}
which is a first-order differential operator on the space
$\mathbb{R}^2\times \mathbb{R}^1$ with coordinates $t , x,$ and $u$, where the coefficients $\tau$ and $\xi$ do not simultaneously vanish.
This operator allows one to construct an
ansatz reducing the original equation \eqref{g1eqVarCoefNonLWaveEq} to an ordinary differential equation.
The conditional invariance criterion \cite{Zhdanov&Tsy&Pop1999,Fushchych&Zhdanov1992,Kun&Pop2009} for
equation~\eqref{g1eqVarCoefNonLWaveEq} to be invariant with respect to
the operator~\eqref{VectorFieldofNonClaSymm} read as
\begin{equation}\label{DeterEq}
\pr^{(2)}Q(L[u])\bigg{|}_{\mathcal{L}\cap
\mathcal{Q}_{(2)}}= 0,
\end{equation}
where $\pr^{(2)}Q$ is the usual second order prolongation~\cite{Olver1986,Ovsiannikov1982} of the
operator~\eqref{VectorFieldofNonClaSymm}, $\mathcal{L}$ is the manifold in the second-order
jet space $J^{(2)}$ determined by the wave equation $L[u] = 0$, and $\mathcal{Q}_{(2)}\subset J^{(2)}$ is the the first prolongation of the invariant surface condition
\begin{equation}\label{AddiCon}
Q[u]:=\tau u_t+\xi u_x-\eta=0.
\end{equation}
The system $\mathcal{Q}_{(2)}$ consists of \eqref{AddiCon} and the equations
obtained by $t-$ and $x-$differentiation of \eqref{AddiCon}.

Below, according to the singular reduction operator theory \cite{Kun&Pop2008}, we first partition the set of
reduction operators of class \eqref{g1eqVarCoefNonLWaveEq} into two subsets, i.e., the singular reduction operator and the regular one.
Then we utilize the two kinds of operators to derive determining equations (overdetermined
system of nonlinear PDEs with respect to the coefficients of the reduction operator \eqref{VectorFieldofNonClaSymm}) from the conditional invariance criterion~\eqref{DeterEq} separately. Solving the two systems we can obtain the final reduction operators. In particular, we will
present a exhausted classification of
the regular operators of class \eqref{g1eqVarCoefNonLWaveEq} by solving the corresponding determining equations.
In general, every Lie symmetry operator is also a reduction operator. Therefore, in this paper we will concentrate on the regular reduction operators
which is inequivalent to Lie symmetry operators, called nontrivial one.

\subsection{Singular reduction operators}

Using the procedure given by Popovych et al. in \cite{Kun&Pop2008},
we can obtain the following assertion.

\begin{proposition}\label{ProSinguOperaOfTeleEq}
A vector field
$Q=\tau(t,x,u)\partial_t+\xi(t,x,u)\partial_x+\eta(t,x,u)\partial_u$
is singular for the differential function $L =
f(x)u_{tt}-(u^nu_x)_x-h(x)u^m$ if and only if~
$\xi^2f(x)=\tau^2u^n$.
\end{proposition}

\begin{proof}
Suppose that $\tau\neq0$. According to the characteristic equation
$\tau u_t+\xi u_x-\eta=0$, we can get
\begin{gather*}
u_t=\frac{\eta}{\tau}-\frac{\xi}{\tau} u_x,\\
u_{tt}=(\frac{\eta}{\tau})_t-(\frac{\xi}{\tau})_tu_x+[(\frac{\eta}{\tau})_u-(\frac{\xi}{\tau})_uu_x](\frac{\eta}{\tau}-\frac{\xi}{\tau}
u_x)
-(\frac{\xi}{\tau})[(\frac{\eta}{\tau})_x+(\frac{\eta}{\tau})_uu_x-(\frac{\xi}{\tau})_xu_x\\-(\frac{\xi}{\tau})_uu_x^2-(\frac{\xi}{\tau})u_{xx}].
\end{gather*}
Substituting the formulas of $u_{tt}$ from above formulaes into $L$,
we obtain a differential function
\begin{gather*}
\tilde
L=[f(x)(\frac{\xi}{\tau})^2-u^n]u_{xx}+f(x)\bigg{\{}(\frac{\eta}{\tau})_t-(\frac{\xi}{\tau})_tu_x
+[(\frac{\eta}{\tau})_u-(\frac{\xi}{\tau})_uu_x](\frac{\eta}{\tau}-\frac{\xi}{\tau}
u_x)
\\-(\frac{\xi}{\tau})[(\frac{\eta}{\tau})_x+(\frac{\eta}{\tau})_uu_x-(\frac{\xi}{\tau})_xu_x
-(\frac{\xi}{\tau})_uu_x^2]\bigg{\}}-nu^{n-1}u_x^2-hu^mu_x.
\end{gather*}
According to the definition 4 of singular
vector field in \cite{Kun&Pop2008}, we have $\ord\tilde L < 2$ if and only if
$f(x)(\frac{\xi}{\tau})^2-u^n=0$.
\end{proof}

Therefore, for any  $f, h, n$ and $m$ with $fu^n>0$  the differential
function $L = f(x)u_{tt}-(u^nu_x)_x-h(x)u^m$ possesses exactly
two set of singular vector fields in the reduced form, namely, $S=
\{ \partial_t +\sqrt{u^n/f}\partial_x+\hat \eta
\partial_u\}$ and $S^*= \{\partial_t
-\sqrt{u^n/f}\partial_x+\hat \eta \partial_u\}$, where $\hat
\eta=\frac{\eta}{\tau}$. Any singular vector field of $L$ is
equivalent to one of the above fields. The singular sets are mapped to each other
by alternating the sign of $x$ and hence one of them can be excluded from the consideration.

\begin{proposition}\label{Poro2SinguOperaOfTeleEq}
For any  variable coefficient nonlinear wave equations in the
form \eqref{g1eqVarCoefNonLWaveEq} the differential function $L =
f(x)u_{tt}-(u^nu_x)_x-h(x)u^m$  possesses exactly one set of
singular vector fields in the reduced form, namely, $S= \{
\partial_t +\sqrt{u^n/f}\partial_x+\hat \eta \partial_u\}$.
\end{proposition}

Thus taking into accountant the conditional invariance criterion for
an equation from class \eqref{g1eqVarCoefNonLWaveEq} and the
operator $\partial_t +\sqrt{u^n/f}\partial_x+\eta \partial_u$, we can
get

\begin{theorem}\label{TheoDESinguOperaOfTeleEq}
Every singular reduction operator of an equation from class
\eqref{g1eqVarCoefNonLWaveEq} is equivalent to an operator of the form
\[
Q=\partial_t +\sqrt{u^n/f(x)}\partial_x+ \eta(t,x,u)
\partial_u,
\]
where the real-valued function $\eta(t,x,u)$ satisfies the
determining equations
\begin{equation}
\begin{array}{ll}
(-1/2nhu^{m-1}-2f\eta_{tu}-2f\eta\eta_{uu})\sqrt{u^n/f}+(3/4f_x^2/f-1/2f_{xx})(u^n/f)^{3/2}\\
+(-1/4n^2\eta^2u^{n-2}-n\eta\eta_uu^{n-1}+1/2n\eta^2u^{n-2})(u^n/f)^{-1/2}-n\eta_xu^{n-1}\\
-1/2n\eta u^{n-1}f_x/f-2\eta_{xu}u^n=0,\\
-\eta_{xx}u^n-n\eta\eta_xu^{n-1}(u^n/f)^{-1/2}+h\eta_uu^m+2f\eta\eta_{tu}+f\eta^2\eta_{uu}\\
-mh\eta u^{m-1}+f\eta_{tt}+(hf_x/f-h_x)u^m\sqrt{u^n/f}=0.
\end{array}
\end{equation}
\end{theorem}

\subsection{Regular reduction operators}

The above investigation of singular reduction operators of nonlinear
wave equation of the form \eqref{g1eqVarCoefNonLWaveEq} shows
that for these equations the regular case of the natural partition of
the corresponding sets of reduction operators is singled out by the
conditions $\xi\neq \pm \sqrt{u^n/f}\tau$. After factorization with
respect to the equivalence relation of vector fields, we obtain the
defining conditions of regular subset of reduction operator:
$\tau=1, \xi\neq \pm \sqrt{u^n/f}$. Hence we have

\begin{proposition}\label{PoroReduOperaOfTeleEq}
For any  variable coefficient nonlinear wave equations in the
form \eqref{g1eqVarCoefNonLWaveEq} the differential function $L =
f(x)u_{tt}-(u^nu_x)_x-h(x)u^m$ possesses exactly one set of
regular vector fields in the reduced form, namely, $S= \{
\partial_t +\hat \xi \partial_x+\hat \eta \partial_u\}$ with $\hat \xi\neq \pm \sqrt{u^n/f}$.
\end{proposition}

Consider the conditional invariance criterion for an equation from
class \eqref{g1eqVarCoefNonLWaveEq} and the operator $\partial_t +
\xi(t,x,u) \partial_x+ \eta(t,x,u) \partial_u$ with $ \xi(t,x,u)\neq \pm \sqrt{u^n/f}$,
we can get the following determining equations for the coefficients $\xi$ and $\eta$:
\begin{equation} \label{PreDEReguReduOpe}
\begin{array}{ll}
\xi_u=0,\quad 2f\xi_t-n\eta u^{n-1}+(2\xi_x+\xi f_x/f)u^n=0,  \\
(2n\xi_x-n\eta_u+n\xi f_x/f)u^{n-1}+(n\eta-n^2\eta)u^{n-2}-\eta_{uu}u^n+f\xi^2\eta_{uu}=0, \\
2f\xi_t\xi_x-2f\xi_t\eta_u-2n\eta_xu^{n-1}-f\xi_{tt}-2f\xi\eta\eta_{uu}-2f\xi\eta_{tu}+(\xi_{xx}-2\eta_{xu})u^n=0,  \\
(\xi hf_x/f-\xi h_x+h\eta_u)u^m+f\eta^2\eta_{uu}+2f\eta\eta_{tu}-2f\xi_t\eta_x-\eta_{xx}u^n+f\eta_{tt}-mh\eta u^{m-1}=0.
\end{array}
\end{equation}
From the first two equations of system \eqref{PreDEReguReduOpe}, we have
\begin{equation*}
\xi=\xi(t,x),\quad \eta=\frac{1}{n}2f\xi_tu^{1-n}+\frac{1}{n}(2\xi_x+\xi\frac{f_x}{f})u.
\end{equation*}
Substituting these expression into the last three equations of system \eqref{PreDEReguReduOpe}, we have the following
assertion.

\begin{theorem}\label{TheoDEReguOperaOfTeleEq}
Every regular reduction operator of an equation from class
\eqref{g1eqVarCoefNonLWaveEq} is equivalent to an operator of the form
\begin{equation}\label{GeneForReguOpe}
Q=\partial_t + \xi(t,x)\partial_x+ \eta(t,x,u)
\partial_u \quad \mbox{with}\quad \eta(t,x,u)=\frac{1}{n}2f\xi_tu^{1-n}+\frac{1}{n}(2\xi_x+\xi\frac{f_x}{f})u,
\end{equation}
where the real-valued function $ \xi(t,x)$ satisfies
the overdetermined system of partial differential equations
\begin{equation} \label{DEReguReduOpe}
\begin{array}{ll}
2(1-n)f\xi_t=0,\quad 2(1-n)f^2\xi^2\xi_t=0,\quad 8(1-n)f^3\xi\xi_t^2=0,  \\
4(1-n)[(f\xi_t)^2-f^2\xi\xi_t(2\xi_x+\xi \frac{f_x}{f})+f^2\xi\xi_{tt}]=0,  \\
\xi_{xx}-2(1+\frac{1}{n})(2\xi_x+\xi \frac{f_x}{f})_x=0, \\
2f\xi_t\xi_x-\frac{2}{n}f\xi_t(2\xi_x+\xi \frac{f_x}{f})-4(f\xi_t)_x-f\xi_{tt}-\frac{2}{n}f\xi(2\xi_x+\xi \frac{f_x}{f})_t-\frac{4}{n}(1-n)(f\xi_t)_x=0,\\
(\xi h\frac{f_x}{f}-\xi h_x+\frac{1}{n}h(1-m)(2\xi_x+\xi \frac{f_x}{f}))u^m+\frac{2}{n}(1-n-m)fh\xi_tu^{m-n}\\
-\frac{8}{n^2}(1-n)f^4(\xi_t)^3u^{1-3n}+\frac{8}{n^2}(1-n)[f^2\xi_t(f\xi_t)_t-f^3\xi_t^2(2\xi_x+\xi \frac{f_x}{f})]u^{1-2n}\\
+[\frac{2}{n}f(f\xi_t)_{tt}-\frac{4}{n}f\xi_t(f\xi_t)_x+\frac{4}{n^2}f^2\xi_t(2\xi_x+\xi \frac{f_x}{f})_t-\frac{2}{n^2}(1-n)f^2\xi_t(2\xi_x+\xi \frac{f_x}{f})^2\\
+\frac{4}{n^2}(1-n)f(f\xi_t)_t(2\xi_x+\xi \frac{f_x}{f})]u^{1-n}+[\frac{2}{n^2}f(2\xi_x+\xi \frac{f_x}{f})(2\xi_x+\xi \frac{f_x}{f})_t\\
+\frac{1}{n}f(2\xi_x+\xi \frac{f_x}{f})_{tt}-\frac{2}{n}f\xi_t(2\xi_x+\xi \frac{f_x}{f})_x-\frac{2}{n}(f\xi_t)_{xx}]u-\frac{1}{n}(2\xi_x+\xi \frac{f_x}{f})_{xx}u^{n+1}=0.
\end{array}
\end{equation}
\end{theorem}

Solving the above system with respect to the coefficient functions $
\xi, f$ and $h$ under the equivalence group
$G_1^{\sim}$ of the class \eqref{g1eqVarCoefNonLWaveEq} which consists of the transformations (see theorem 3 and 4 in \cite{Huang&Yang&Zhou2010a} for more details): for $n\neq-1$
\[
\begin{array}{ll}
\tilde t =\epsilon_1 t +\epsilon_2,\quad \tilde x=\frac{\epsilon_3x+\epsilon_4}{\epsilon_5x+\epsilon_6}=: X(x),\quad
\tilde u =\epsilon_7 X_x^{\frac{1}{2n+2}} u ,\\
\tilde f =\epsilon_1^2\epsilon_7^nX_x^{-\frac{3n+4}{2n+2}}f,\quad \tilde h =
\epsilon_7^{-m+n+1}X_x^{-\frac{m+3n+3}{2n+2}}h,\quad \tilde n= n ,\quad
\tilde m=m,
\end{array}
\]
where $\epsilon_j~(j=1,\ldots, 7)$ are
arbitrary constants,
$\epsilon_1\epsilon_7\neq 0, \epsilon_3\epsilon_6-\epsilon_4\epsilon_5=\pm 1$ and for $n=-1$
\[
\begin{array}{ll}
\tilde t =\epsilon_1 t +\epsilon_2,\quad \tilde x=\epsilon_3x+\epsilon_4,\quad
\tilde u =\epsilon_5 e^{\epsilon_6x} u ,\\
\tilde f =\epsilon_1^2\epsilon_3^{-2}\epsilon_5^{-1}e^{-\epsilon_6x}f,\quad \tilde h =
\epsilon_3^{-2}\epsilon_5^{-m}e^{-m\epsilon_6x}h,\quad \tilde n= n ,\quad
\tilde m=m,
\end{array}
\]
where $\epsilon_j~(j=1,\ldots, 6)$ are
arbitrary constants,
$\epsilon_1\epsilon_3\epsilon_5\neq 0$; we can get a classification of regular reduction
operator for the class \eqref{g1eqVarCoefNonLWaveEq}. It is easy to know that
some of the regular reduction operator are equivalent to Lie symmetry operators,
while some of are nontrivial. Below, we give a detailed investigations for these cases.

In fact, the first three equations of system \eqref{DEReguReduOpe} implies there are two cases should be considered: $n\neq 1$ or not. (It should be noted that $\xi = 0$
should be exclude from the consideration because it leads to $\eta=0$). \\

{\bf Case 1:} $\; n\neq 1$. In this case, we have
$\xi_t=0$. Thus system \eqref{DEReguReduOpe} can be reduced to
\begin{equation}\label{ReguDEsystem1}
\begin{array}{ll}
(3n+4)\xi_{xx}+2(n+1)(\xi\frac{f_x}{f})_x=0,\\
(\xi h\frac{f_x}{f}-\xi h_x+\frac{1}{n}(1-m) h (2\xi_x+\xi \frac{f_x}{f})) u^m-\frac{1}{n}(2\xi_x+\xi \frac{f_x}{f})_{xx} u^{n+1}=0.
\end{array}
\end{equation}
Thus, there are two cases should be considered: $m\ne n+1$ or not.

{\it Case 1.1:} For $m\neq n+1$, from the second equation of \eqref{ReguDEsystem1} we obtain
\begin{equation} \label{ReguDEsystem1.1}
\xi h\frac{f_x}{f}-\xi h_x+\frac{1}{n}(1-m) h (2\xi_x+\xi\frac{f_x}{f})=0, \quad
\big(2\xi_x+\xi\frac{f_x}{f}\big)_{xx}=0.
\end{equation}
The first equation of \eqref{ReguDEsystem1} suggests that
$(3n+4)\xi_x+2(n+1)\xi\frac{f_x}{f}$ is independent of the variable $x$,
so there exists a constant $r$ such that
$(3n+4)\xi_x+2(n+1)\xi\frac{f_x}{f}=nr$.
The second equation of \eqref{ReguDEsystem1.1} suggests that there exist two constants
$a$ and $b$ such that
$2\xi_x+\xi\frac{f_x}{f}=nax+nb$.
By solving the last two equations we obtain
\[
\xi_x=2(n+1)(ax+b)-r,\quad
\xi\frac{f_x}{f}=2r-(3n+4)(ax+b),
\]
which together with the first equation of  \eqref{ReguDEsystem1.1} imply
\begin{equation}\label{1RuguOperaOfEq}
\begin{array}{ll}
\xi=a(n+1)x^2+[2b(n+1)-r]x+s, \\
\displaystyle f(x)=\exp \big(\int \dfrac{2r-(3n+4)(ax+b)}{a(n+1)x^2+[2b(n+1)-r]x+s}\, {\rm d}x  \big), \\
\displaystyle h(x)=\exp \big(\int \dfrac{2r-(m+3n+3)(ax+b)}{a(n+1)x^2+[2b(n+1)-r]x+s}\, {\rm d}x  \big),
\end{array}
\end{equation}
where $a,b,r,s$ are arbitrary constants. Thus, the corresponding regular reduction operator has the form
\[
Q=\partial_t + [a(n+1)x^2+(2b(n+1)-r)x+s]\partial_x+ (ax+b)u\partial_u,
\]
which is equivalent to Lie symmetry operator.

{\it Case 1.2:} $m= n+1$.
In this case, system \eqref{ReguDEsystem1} can be rewritten as
\begin{equation}\label{ReguDEsystem1.2}
\begin{array}{ll}
(3n+4)\xi_{xx}+2(n+1)(\xi\frac{f_x}{f})_x=0,\\
\xi h_x+2h \xi_x+\frac{1}{n}(2\xi_x+\xi\frac{f_x}{f})_{xx}=0.
\end{array}
\end{equation}
Integrating these two equations with respect to functions $f(x)$ and $g(x)$, we can obtain
\[
\displaystyle f(x)=\vert\, \xi \vert ^{-\frac{3n+4}{2n+2}}\exp \big(r\int \frac{1}{\xi}\,{\rm d} x\big), \quad
\displaystyle h(x)=\dfrac{\xi_x^2-2\xi\xi_{xx}-p}{4(n+1)\xi^2}.
\]
where $p, r$ are arbitrary constants, $\xi$ is an arbitrary smooth function and $n\neq -1$. In addition, $\eta=\frac{1}{n}(2\xi_x+\xi\frac{f_x}{f})u=(\frac{r}{n}+\frac{\xi_x}{2n+2})u$.
Thus, we have a nontrivial regular reduction operator
\begin{equation}\label{2RuguOperaOfEq}
Q=\partial_t + \xi(x)\partial_x+[(\frac{r}{n}+\frac{\xi_x}{2n+2})u]\partial_u, \quad n\neq-1.
\end{equation}
It should be noted that for $n=-1$ the reduction operator is also equivalent to Lie symmetry operator.\\

{\bf Case 2:} $\; n=1$. In this case, we have $\eta=2f\xi_t+(2\xi_x+\xi\frac{f_x}{f})u$. Thus, system \eqref{DEReguReduOpe} can be reduced to
\begin{equation}\label{ReguDEsystem2}
\begin{array}{ll}
(7\xi_x+4\xi\frac{f_x}{f})_x=0, \quad
2\big[\xi_x+2(\xi+1)\frac{f_x}{f}\big]\xi_t+4(\xi +1)\xi_{tx}+\xi_{tt}=0, \\
\big[2f(2\xi_x+\xi\frac{f_x}{f})(2\xi_x+\xi\frac{f_x}{f})_t+f(2\xi_x+\xi\frac{f_x}{f})_{tt}-2f\xi_t(2\xi_x+\xi\frac{f_x}{f})_x-2(f\xi_t)_{xx}\big]u\\-(2\xi_x+\xi\frac{f_x}{f})_{xx}u^2
+2f^2(2\xi_t\xi_{tx}+\xi_{ttt})
+\big[ \xi h\frac{f_x}{f}-\xi h_x+(1-m)h(2\xi_x+\xi\frac{f_x}{f}) \big]u^m\\
-2mhf\xi_t u^{m-1} =0.
\end{array}
\end{equation}
After some brief analysis, we find that there are five cases should be considered.\\

{\it Case 2.1:} $m=0$. In this case, the third equation of \eqref{ReguDEsystem2} implies
\begin{equation}\label{ReguDEsystem3}
\begin{array}{ll}
2h\xi\frac{f_x}{f}-h_x\xi+2h\xi_x+4f^2\xi_t\xi_{tx}+2f^2\xi_{ttt}=0, \\
2f(2\xi_x+\xi\frac{f_x}{f})(2\xi_x+\xi\frac{f_x}{f})_t+f(2\xi_x+\xi\frac{f_x}{f})_{tt}-2f\xi_t(2\xi_x+\xi\frac{f_x}{f})_x-2(f\xi_t)_{xx}=0, \\
(2\xi_x+\xi\frac{f_x}{f})_{xx}=0.
\end{array}
\end{equation}
From the last equation of system \eqref{ReguDEsystem3} we can know that there exist two functions $a(t)$ and $b(t)$
such that $2\xi_x+\xi\frac{f_x}{f}=a(t)x+b(t)$. On the other hand,
the first equation of \eqref{ReguDEsystem2} implies there exists a function $c(t)$
such that $7\xi_x+4\xi\frac{f_x}{f}=c(t)$.
Solving the last two equations gives
\[
\xi_x=4a(t)x+4b(t)-c(t), \quad \xi\frac{f_x}{f}=-7a(t)x-7b(t)+2c(t),
\]
from which we can get
\begin{equation}\label{ReguDEsystem3.1}
\begin{array}{ll}
\xi=2a(t)x^2+4b(t)x-c(t)x+d(t), \\
\displaystyle f(x)=\exp\Big(\int \dfrac{-7a(t)x-7b(t)+2c(t)}
{2a(t)x^2+4b(t)x-c(t)x+d(t)}\,{\rm d}\, x\Big).
\end{array}
\end{equation}
where $d(t)$ is an arbitrary functions.
Since $\frac{f_x}{f}$ is independent of $t$, we see that
\[
\Big[ \dfrac{-7a(t)x-7b(t)+2c(t)}{2a(t)x^2+4b(t)x-c(t)x+d(t)}\Big]_t=0,
\]
which leads to
\begin{equation} \label{ReguDEsystem3.2}
\begin{cases}
14\big[ a(t)b\,'(t)-a\,'(t)b(t)\big]+3\big[a\,'(t)c(t)-a(t)c\,'(t)\big]=0, \\
\big[ b(t)c\,'(t)-b\,'(t)c(t) \big]+7\big[ a(t)d\,'(t)-a\,'(t)d(t) \big]=0, \\
2\big[ c\,'(t)d(t)-c(t)d\,'(t)\big]+7\big[ b(t)d\,'(t)-b\,'(t)d(t)\big]=0.
\end{cases}
\end{equation}

Now, we multiply both sides of the second equation of
\eqref{ReguDEsystem2} by $\xi$ and substitute \eqref{ReguDEsystem3.1}
into it, then simplify the equation and compare the coefficient of
$x^{i}$($i=0,1,\ldots,5$)
to obtain
\begin{equation} \label{ReguDEsystem3.3}
\begin{cases}
a\,'(t)=0,\\
a^2[-4b\,'(t)+c\,'(t)]=0,\\
a[8c(t)c\,'(t)+20ad\,'(t)+c\,''(t)-4b\,''(t)-32c(t)b\,'(t)+112b(t)b\,'(t)-28b(t)c\,'(t)]=0,\\
-4c\,''(t)b(t)+32ac(t)d\,'(t)-80ab\,'(t)+20ac\,'(t)-4b\,''(t)c(t)-120ab(t)d\,'(t)\\
+c\,''(t)c(t)+2d\,''(t)a+2c^2(t)c\,'(t)-12b(t)c(t)c\,'(t)+16b(t)^2c\,'(t)-8c^2(t)b\,'(t)\\
+48b(t)c(t)b\,'(t)+16b\,''(t)b(t)-64b^2(t)b\,'(t)-16ad(t)b\,'(t)+4ad(t)c\,'(t)=0,\\
[-20ad\,'(t)+48b(t)b\,'(t)+2c(t)c\,'(t)-12b(t)c\,'(t)-8c(t)b\,'(t)+4b\,''(t)-c\,''(t)]d(t)\\
-80b^2(t)d\,'(t)+4d\,''(t)b(t)-d\,''(t)c(t)-4c(t)c\,'(t)+44b(t)c(t)d\,'(t)-28ad\,'(t) \\
-48b(t)b\,'(t)-6c^2(t)d\,'(t)+12b(t)c\,'(t)+16c(t)b\,'(t)=0,\\
[16b\,'(t)-4c\,'(t)]d^2(t)+[-20b(t)d\,'(t)-4c\,'(t)+6c(t)d\,'(t)+d\,''(t)+16b\,'(t)]d(t)\\
+8c(t)d\,'(t)-28b(t)d\,'(t)=0.
\end{cases}
\end{equation}

Note that $\xi$ is assumed not to be identical with zero,
after some simple but lengthy computations, we find that systems \eqref{ReguDEsystem3.2} and \eqref{ReguDEsystem3.3}
can be reduced to:
\begin{equation} \label{ReguDEsystem3.4}
a\,'(t)=0,\quad b\,'(t)=0,\quad c\,'(t)=0,\quad d\,'(t)=0
\end{equation}
or
\begin{equation} \label{ReguDEsystem3.5}
a=0, \quad c(t)=4b(t),\quad d(t)=qb(t), \quad qb\,''(t)+4qb(t)b\,'(t)+4b\,'(t)=0
\end{equation}
or
\begin{equation} \label{ReguDEsystem3.6}
a=0, \quad 2c(t)=7b(t),\quad b\,''(t)=-3b(t)b\,'(t), \quad
b(t)d\,'(t)+2b\,'(t)(d(t)+1)+d\,''(t)=0
\end{equation}
or
\begin{equation} \label{ReguDEsystem3.7}
a=0, \quad c(t)=3b(t),\quad d=qb(t), \quad b\,''(t)+2b(t)b\,'(t)=0,
\end{equation}
where $q$ is an arbitrary constant.

{\it Case 2.1a:}  If system \eqref{ReguDEsystem3.4} is satisfied, then $\xi_t=0$,
the second equation of \eqref{ReguDEsystem3} is an identity.
The expression \eqref{ReguDEsystem3.1} can be rewritten as
\begin{equation} \label{ReguDEsystem3.4.1}
\begin{array}{ll}
\xi=2ax^2+4bx-cx+d, \\
\displaystyle f(x)=\exp\Big(\int \dfrac{-7ax-7b+2c}
{2ax^2+4bx-cx+d}\,{\rm d}\, x\Big),
\end{array}
\end{equation}
where $a, b, c$ and $d$ are arbitrary constants.
The first equation of \eqref{ReguDEsystem3} is reduced to
\[
\dfrac{h_x}{h}=\dfrac{2\big(\xi\frac{f_x}{f}+\xi_x\big)}{\xi}.
\]
Substitute the expression of $\xi$ and $f(x)$ into it and
integrate both sides to obtain
\[
\displaystyle h(x)=\exp\Big(\int \dfrac{-6ax-6b+2c}
{2ax^2+4bx-cx+d}\,{\rm d}\, x\Big).
\]
In addition, $\eta=2f\xi_t+(2\xi_x+\xi\frac{f_x}{f})u=(ax+b)u$.
Therefore, we have
\begin{equation}\label{3RuguOperaOfEq}
\begin{array}{ll}
\xi=2ax^2+4bx-cx+d,  \\
\eta=(ax+b)u, \\
\displaystyle f(x)=\exp\Big(\int \dfrac{-7ax-7b+2c}
{2ax^2+4bx-cx+d}\,{\rm d}\, x\Big),\\
\displaystyle h(x)=\exp\Big(\int \dfrac{-6ax-6b+2c}
{2ax^2+4bx-cx+d}\,{\rm d}\, x\Big).
\end{array}
\end{equation}
where $a, b, c, d$ are arbitrary constants. Thus, the corresponding regular reduction operator has the form
\[
Q=\partial_t + (2ax^2+4bx-cx+d)\partial_x+ (ax+b)u\partial_u,
\]
which is equivalent to Lie symmetry operator.

{\it Case 2.1b:}  If system \eqref{ReguDEsystem3.5} is satisfied, then the
expression \eqref{ReguDEsystem3.1} can be rewritten as
\begin{equation}\label{ReguDEsystem3.5.1}
\begin{array}{ll}
\xi=qb(t),\quad
\displaystyle f(x)=\exp\big(\frac{x}
{q}\big).
\end{array}
\end{equation}
Hence, $\xi_x=0$, $k=b(t)$.
Substituting these formulaes into the second equation of
\eqref{ReguDEsystem3} we obtain
\[
qb\,''(t)+2qb(t)b\,'(t)-2b\,'(t)=0.
\]
Combine it with the fourth equation
of \eqref{ReguDEsystem3.5} to get $b\,'(t)=0$.
Hence $a(t),b(t),c(t),d(t)$ satisfy system \eqref{ReguDEsystem3.4}, and the solution is included in the case 2.1a.

{\it Case 2.1c:}  If system \eqref{ReguDEsystem3.6} is satisfied, then the
expression \eqref{ReguDEsystem3.1} can be rewritten as
\begin{equation}\label{ReguDEsystem3.6.1}
\begin{array}{ll}
\xi=\frac{1}{2}b(t)x+d(t), \quad
f(x)=1 \mod G_1^{\sim}.
\end{array}
\end{equation}
Substitute it into the second equation of
\eqref{ReguDEsystem3} to obtain
$2b(t)b\,'(t)+b\,''(t)=0$.
Combine it with the third equation
of \eqref{ReguDEsystem3.6} to get $b\,'(t)=0$.
Substitute it into the fourth equation of
\eqref{ReguDEsystem3.6} to obtain
$bd\,'(t)+d\,''(t)=0$,
which implies $d(t)=\gamma_1e^{-bt}+\gamma_0$,
where $\gamma_1$ and $\gamma_0$ are arbitrary constants.
Therefore $\xi=\frac{b}{2}x+\gamma_1e^{-bt}+\gamma_0$.
Substitute it into the first equation of
\eqref{ReguDEsystem3} to obtain
\[
2\gamma_1(h_x+2b^3)e^{-bt}+bxh_x+2\gamma_0h_x-2bh=0.
\]
Since $h, \gamma_1, \gamma_0$ are independent of $t$,
the preceding equation suggests that
\[
\begin{array}{ll}
2\gamma_1(h_x+2b^3)=0, \quad
bxh_x+2\gamma_0h_x-2bh=0,
\end{array}
\]
which leads
to $b=0$. Therefore, we have
\[
\xi=d_1t+d_0,\quad \eta=2d_0,\quad f=1,\quad h=h_0 \mod G^{\sim},
\]
where $d_1, d_0, h_0$ are constants.  Thus, the corresponding regular reduction operator has the form
\[
Q=\partial_t + (d_1t+d_0)\partial_x+ 2d_1u\partial_u,
\]
which is equivalent to Lie symmetry operator.

{\it Case 2.1d:}  If system \eqref{ReguDEsystem3.7} is satisfied, then the
expression \eqref{ReguDEsystem3.1} can be rewritten as
\begin{equation}\label{ReguDEsystem3.7.1}
\begin{array}{ll}
\xi=b(t)(x+q), \quad
f(x)=\dfrac{1}{x+q} \mod G_1^{\sim}.
\end{array}
\end{equation}
Substitute it into the second equation of
\eqref{ReguDEsystem3} to obtain
$2b(t)b\,'(t)+b\,''(t)=0$,
which is equivalent to the fourth equation
of \eqref{ReguDEsystem3.7}, and which
leads to
$2b\,'^2(t)+b\,'''(t)=-2b(t)b\,''(t)$.
Substitute \eqref{ReguDEsystem3.7.1}
into the first equation of
\eqref{ReguDEsystem3} to obtain
\[
h_x=2f^2[2b\,'^{\,2}(t)+b\,'''(t)]/b(t)
=2f^2[-2b(t)b\,''(t)]/b(t)=-4f^2b\,''(t).
\]
Since $h$ and $f$ are independent of $t$, there is a constant
$r$ such that $b\,''(t)=r$.
It follows that there exist constants $s$ and $w$
such that $b(t)=rt^2/2+st+w$. Substitute it into the fourth equation of
 \eqref{ReguDEsystem3.7} to obtain
\[
r^2t^3+3rst^2+2(wr+s^2)t+2ws+r=0.
\]
Then $r=0$ and $s=0$. Hence $b(t)=w$, $\xi_t=0$, the solution is included in the case 2.1a.\\

{\it Case 2.2:}  When $m=1$, the third equation of \eqref{ReguDEsystem2} implies
\begin{equation}\label{ReguDEsystem4}
\begin{array}{ll}
\xi h \frac{f_x}{f}-\xi h_x+f(2\xi_x+\xi\frac{f_x}{f})_{tt}+2f(2\xi_x+\xi\frac{f_x}{f})(2\xi_x+\xi\frac{f_x}{f})_t\\
-2f\xi_t(2\xi_x+\xi\frac{f_x}{f})_x-2(f\xi_t)_{xx}=0, \\
 2f\xi_t\xi_{tx}+f\xi_{ttt}-h\xi_t = 0, \\
(2\xi_x+\xi\frac{f_x}{f})_{xx}=0.
\end{array}
\end{equation}
Similar to the case of $m=0$, from the third equation of \eqref{ReguDEsystem4}
and the first two equations of
\eqref{ReguDEsystem2} we get the expression of $\xi$ and $f(x)$
as stated in \eqref{ReguDEsystem3.1},
where $a(t),b(t),c(t),d(t)$ satisfy the condition
 \eqref{ReguDEsystem3.4} or \eqref{ReguDEsystem3.5} or
 \eqref{ReguDEsystem3.6} or \eqref{ReguDEsystem3.7}.

{\it Case 2.2a:}  If system \eqref{ReguDEsystem3.4} is satisfied, then $\xi_t=0$,
the second equation of \eqref{ReguDEsystem4} is an identity.
The expression \eqref{ReguDEsystem3.1} can be rewritten as \eqref{ReguDEsystem3.4.1}.
The first equation of \eqref{ReguDEsystem4} is reduced to
$h_x/h=f_x/f$, which leads to $h(x)=\epsilon f(x) ~(\epsilon=\pm 1) \mod G_{1}^{\sim}$.
In addition, $\eta=2f\xi_t+(2\xi_x+\xi\frac{f_x}{f})u=(ax+b)u$. Thus, we have
\begin{equation}\label{4RuguOperaOfEq}
\begin{array}{ll}
\xi=2ax^2+4bx-cx+d,  \\
\eta=(ax+b)u, \\
\displaystyle f(x)=\exp\Big(\int \dfrac{-7ax-7b+2c}
{2ax^2+4bx-cx+d}\,{\rm d}\, x\Big),\\
\displaystyle h(x)=\epsilon f(x).
\end{array}
\end{equation}
where $a, b, c, d$ are arbitrary constants and $\epsilon=\pm 1$. Thus, the corresponding regular reduction operator has the form
\[
Q=\partial_t + (2ax^2+4bx-cx+d)\partial_x+ (ax+b)u\partial_u,
\]
which is equivalent to Lie symmetry operator.

{\it Case 2.2b:}  If system \eqref{ReguDEsystem3.5} is satisfied, then the
expression \eqref{ReguDEsystem3.1} can be rewritten as
\eqref{ReguDEsystem3.5.1}. Hence, $\xi_x=0$, $k=b(t)$.
If $b\,'(t)=0$, then $a(t),b(t),c(t),d(t)$ satisfy system
 \eqref{ReguDEsystem3.4}, and the solution
 is included in the case 2.2a.
 We suppose that $b\,'(t)\ne 0$. From the second equation of
\eqref{ReguDEsystem4} we see that $h=f\xi_{ttt}/\xi_t$.
Substitute it into the first equation
of \eqref{ReguDEsystem4} to get $fk_{tt}+2fkk_t-2f_{xx}\xi_t=0$.
Further it can be reduced to
$qb\,''(t)+2qb(t)b\,'(t)-2b\,'(t)=0$.
Combine it with the fourth equation of \eqref{ReguDEsystem3.5}
to get $b(t)=-3/q$ which is contradict to the hypothesis
$b\,'(t)\ne 0$.

{\it Case 2.2c:}  If system \eqref{ReguDEsystem3.6} is satisfied, then the
expression \eqref{ReguDEsystem3.1} can be rewritten as
\eqref{ReguDEsystem3.6.1}.
If $b\,'(t)=d\,'(t)=0$, then $a(t),b(t),c(t),d(t)$ satisfy  both systems
 \eqref{ReguDEsystem3.4} and  \eqref{ReguDEsystem3.6}, and the solution
 is included in the case 2.2a.
 We suppose that $b\,'^2(t)+d\,'^2(t)\ne 0$.
Substitute \eqref{ReguDEsystem3.6.1} into the first equation of
\eqref{ReguDEsystem4} to obtain
\begin{equation} \label{ReguDEsystem4.1}
h_x=\dfrac{2[b\,''(t)+2b(t)b\,''(t)]}{b(t)x+2d(t)}
\end{equation}
Substitute \eqref{ReguDEsystem3.6.1} into the second equation of
\eqref{ReguDEsystem4} to obtain
\begin{equation} \label{ReguDEsystem4.2}
h(x)=b\,'(t)+\dfrac{b\,'''(t)x+2d\,'''(t)}{b\,'(t)x+2d\,'(t)}
\end{equation}
Then
\[
h_x=\dfrac{2[b\,'''(t)d\,'(t)-b\,'(t)d\,'''(t)]}{[b\,'(t)x+2d\,'(t)]^2}.
\]
Substituting it into \eqref{ReguDEsystem4.1} yields
\begin{equation*}
\dfrac{b\,'''(t)d\,'(t)-b\,'(t)d\,'''(t)}{\big[b\,'(t)x+2d\,'(t)\big]^2}=\dfrac{b\,''(t)+2b(t)b\,'(t)}{b(t)x+2d(t)}.
\end{equation*}
Compare the coefficient of $x^2$ to obtain $b\,'^2(t)\big[b\,''(t)+2b(t)b\,'(t)\big]=0$.
Substitute the third equation of \eqref{ReguDEsystem3.6} into it to obtain
$b(t)b\,'^3(t)=0$, hence $b\,'(t)=0$.
Thus the fourth equation of \eqref{ReguDEsystem3.6} can be reduced to
$bd\,'(t)+d\,''(t)=0$.
Solving this linear ordinary differential equation gives
$d(t)=\gamma_1e^{-bt}+\gamma_0$, where $\gamma_1$ and $\gamma_0$ are two arbitrary constants.
Therefore the expressions \eqref{ReguDEsystem3.6.1} and \eqref{ReguDEsystem4.2}
can be rewritten as
\begin{equation*}
\begin{array}{ll}
\xi=\frac{1}{2}bx+\gamma_1e^{-bt}+\gamma_0, \quad f(x)=1, \quad h(x)=b^2 \mod ~G_{1}^{\sim}.
\end{array}
\end{equation*}
System \eqref{ReguDEsystem4} is verified to be true.
In addition, $\eta=2f\xi_t+(2\xi_x+\xi\frac{f_x}{f})u=bu-2\gamma_1be^{-bt}$. Therefore, we have
\begin{equation}\label{5RuguOperaOfEq}
\begin{array}{ll}
\xi=\frac{1}{2}bx+\gamma_1e^{-bt}+\gamma_0, \quad
\eta=bu-2\gamma_1be^{-bt},\\
f(x)=1, \quad  h(x)=b^2,
\end{array}
\end{equation}
where $b, \gamma_1, \gamma_0$ are arbitrary constants. Thus, we have a nontrivial regular reduction operator
\begin{equation}\label{3RuguOperaOfEq}
Q=\partial_t + (\frac{1}{2}bx+\gamma_1e^{-bt}+\gamma_0)\partial_x+(bu-2\gamma_1be^{-bt})\partial_u.
\end{equation}

{\it Case 2.2d:} If system \eqref{ReguDEsystem3.7} is satisfied, then the
expression \eqref{ReguDEsystem3.1} can be rewritten as
\eqref{ReguDEsystem3.7.1}. Substitute it into the first equation
of \eqref{ReguDEsystem4} to obtain
\[
b(t)(x+q)[h+(x+q)h_x]=[b\,''(t)+2b(t)b\,'(t)].
\]
Substitute the fourth equation of \eqref{ReguDEsystem3.7} into it to
get $b(t)(x+q)[h+(x+q)h_x]=0$. It follows that $h(x)=r/(x+q)$,
where $r$ is a nonzero constant.
Substitute it and \eqref{ReguDEsystem3.7.1} into the second equation
of \eqref{ReguDEsystem4} to obtain
$2b\,'^2(t)+b\,'''(t)-rb\,'(t)=0$.
From the fourth equation of \eqref{ReguDEsystem3.7}, we find
$b\,'''(t)=4b^2(t)b\,'(t)-2b\,'^2(t)$. Substitute it into the preceding
equation to get $b\,'(t)[4b^2(t)-r]=0$, which leads to $b\,'(t)=0$. Then
$a(t),b(t),c(t),d(t)$ satisfy system
 \eqref{ReguDEsystem3.4}, and the solution
 is included in the case 2.2a.\\

{\it Case 2.3:}  When $m=2$, system \eqref{ReguDEsystem2} implies
\begin{equation}\label{ReguDEsystem5}
\begin{array}{ll}
(7\xi_x+4\xi\frac{f_x}{f})_x=0, \\
2(\xi_x+2\xi\frac{f_x}{f})\xi_t+4\xi_t\frac{f_x}{f}+4\xi\xi_{tx}+4\xi_{tx}+\xi_{tt}=0, \\
2\xi_t\xi_{tx}+\xi_{ttt}=0, \\
2f(2\xi_x+\xi\frac{f_x}{f})(2\xi_x+\xi\frac{f_x}{f})_t+f(2\xi_x+\xi\frac{f_x}{f})_{tt}-4hf\xi_t-2f\xi_t(2\xi_x+\xi\frac{f_x}{f})_x-2(f\xi_t)_{xx}=0, \\
 \xi h_x+2h\xi_x+(2\xi_x+\xi\frac{f_x}{f})_{xx} = 0.
\end{array}
\end{equation}
From the first and the last equation of system \eqref{ReguDEsystem5}, we can get
\[
\displaystyle f(x)=|\xi|^{-7/4}\exp(\alpha(t)\int\frac{{\rm d}x}{\xi}),\quad \displaystyle h(x)=\frac{\xi_x^2-2\xi\xi_{xx}+q}{8\xi^2},
\]
where $\alpha(t)$ is an arbitrary function, $q$ is a constant. Substituting these expressions into the rest equations of system \eqref{ReguDEsystem5}, we can see that
$\xi(t,x)$ and $\alpha(t)$ satisfy the overdetermined system of partial differential equations
\begin{equation}\label{7RuguOperaOfEq}
\begin{array}{ll}
2\xi_t\xi_{tx}+\xi_{ttt}=0, \\
\xi_{tt}-3\xi_{tx}-5\xi_t\xi_x+4\xi\xi_{tx}+4\alpha\xi_t+4\alpha_t=0, \\
2\xi^2(\frac{1}{4}\xi_x+\alpha)_{tt}+2\xi^2[(\frac{1}{4}\xi_x+\alpha)^2]_{t}-\xi^2\xi_t\xi_{xx}-\xi_t(\xi_x^2-2\xi\xi_{xx}+q)\\
-4[(\alpha_t-\frac{3}{4}\xi_{tx})(\alpha-\frac{7}{4}\xi_{x})\xi-\frac{3}{4}\xi^2\xi_{txx}]=0.
\end{array}
\end{equation}
In addition, we have
\[
\eta=2f\xi_t+(2\xi_x+\xi\frac{f_x}{f})u=2\xi_t|\xi|^{-7/4}\exp(\alpha(t)\int\frac{{\rm d}x}{\xi})+[\frac{1}{4}\xi_x+\alpha(t)]u.
\]
Thus, we have a nontrivial regular reduction operator
\begin{equation}\label{3RuguOperaOfEq}
Q=\partial_t + \xi(t,x)\partial_x+\{2\xi_t|\xi|^{-7/4}\exp(\alpha(t)\int\frac{{\rm d}x}{\xi})+[\frac{1}{4}\xi_x+\alpha(t)]u\}\partial_u,
\end{equation}
where $\xi(t,x)$ and $\alpha(t)$ satisfy the overdetermined system of partial differential equations \eqref{7RuguOperaOfEq}.

In particular, if $\xi_t=0$, from system \eqref{ReguDEsystem5} we can obtain
\begin{equation}\label{6RuguOperaOfEq}
\begin{array}{ll}
\xi=\xi(x), \quad
\eta=\frac{1}{4}(\xi_x+a)u,\\
\displaystyle f(x)=|\xi|^{-7/4}\exp(\frac{a}{4}\int\frac{{\rm d}x}{\xi}),\quad
\displaystyle h(x)=\frac{\xi_x^2-2\xi\xi_{xx}+q}{8\xi^2}.
\end{array}
\end{equation}
where $a, q$ are arbitrary constants. Thus, we have a nontrivial regular reduction operator
\begin{equation}\label{4RuguOperaOfEq}
Q=\partial_t + \xi(x)\partial_x+(\frac{1}{4}\xi_x+a)u\partial_u,
\end{equation}
which is equivalent to operator \eqref{2RuguOperaOfEq} with $n=1$. Therefore, this special case can be included in case 1.2 and we can impose an additional constraint
$\xi_t\neq 0$ on the regular reduction operator \eqref{3RuguOperaOfEq}.\\

{\it Case 2.4:}  When $m=3$, the third equation of \eqref{ReguDEsystem2} implies
\begin{equation}\label{ReguDEsystem6}
\begin{array}{ll}
2\xi_t\xi_{tx}+\xi_{ttt}=0, \\
2f(2\xi_x+\xi\frac{f_x}{f})(2\xi_x+\xi\frac{f_x}{f})_t+f(2\xi_x+\xi\frac{f_x}{f})_{tt}-2f\xi_t(2\xi_x+\xi\frac{f_x}{f})_x-2(f\xi_t)_{xx}=0, \\
6hf\xi_t+(2\xi_x+\xi\frac{f_x}{f})_{xx} = 0, \\
\xi h\frac{f_x}{f}-\xi h_x-2h(2\xi_x+\xi\frac{f_x}{f})=0,
\end{array}
\end{equation}
the fourth equation of which can be rewritten as
\[
\frac{\xi_x}{\xi}=-\frac{1}{4}\big(\frac{f_x}{f}+\frac{h_x}{h}\big).
\]
Since $f$ and $h$ are independent of $t$, integrate
both sides of the preceding equation to obtain
$\xi=r(t)|fh|^{-1/4}$,
where $r(t)$ is a function of $t$.
Substituting it into the first equation of
\eqref{ReguDEsystem6} yields the fact that $r\,'''(t)=q(x)r\,'^{\,2}(t)$, where
$q(x)=-2(|fh|^{-1/4})_x$. It follows that
$r\,'(t)=0$ or $q\,'(x)=0, ~r\,'''(t)=q r\,'^{\,2}(t)$.

{\it Case 2.4a:}  If $r\,'(t)=0$, then $\xi=r|fh|^{-1/4}$ ($r=\const$), $\xi_t=0$, $g_t=0$, the second equation
of \eqref{ReguDEsystem6} is an identity, so is the second equation
of \eqref{ReguDEsystem2}. The third equation of \eqref{ReguDEsystem6} reduces to
$\big(2\xi_x+\xi\frac{f_x}{f}\big)_{xx}=0$. Combine it with the first equation
of \eqref{ReguDEsystem2}, and use a progress similar to the case $m=0$ (i.e. 2.1a), we get
the expression of $\xi$ and $f(x)$ as stated in  \eqref{ReguDEsystem3.4.1},
where $a$, $b$, $c$, $d$ are constants.
From $\xi=r|fh|^{-1/4}$, we see that
$\displaystyle h(x)=\pm\frac{1}{f}\big(\frac{r}{\xi}\big)^{\,4}$,
where $r\xi> 0$.
In addition, $\eta=2f\xi_t+(2\xi_x+\xi\frac{f_x}{f})u=(ax+b)u$.
Thus, we have
\begin{equation}\label{8RuguOperaOfEq}
\begin{array}{ll}
\xi=2ax^2+4bx-cx+d,  \\
\eta=(ax+b)u, \\
\displaystyle f(x)=\exp\Big(\int \dfrac{-7ax-7b+2c}
{2ax^2+4bx-cx+d}\,{\rm d}\, x\Big),\\
\displaystyle h(x)=\pm\frac{1}{f}\big(\frac{r}{\xi}\big)^{\,4}.
\end{array}
\end{equation}
where $a, b, c, d, r$ are arbitrary constants. Thus, the corresponding regular reduction operator has the form
\[
Q=\partial_t + (2ax^2+4bx-cx+d)\partial_x+ (ax+b)u\partial_u,
\]
which is equivalent to Lie symmetry operator.

{\it Case 2.4b:}  If $ q\,'(x)=0, r\,'''(t)=q r\,'^{\,2}(t)$,
then $(|fh|^{-1/4})_x=-\frac{1}{2}q$.
Integration both sides of it gives $|fh|^{-1/4}=-\frac{1}{2}qx+s$,
where $s$ is a constant.
Therefore $\xi=r(t)|fh|^{-1/4}=r(t)(-\frac{1}{2}qx+s)$.
From the first equation of \eqref{ReguDEsystem2} we see that
$(2\xi_x+\xi\frac{f_x}{f})_x=\frac{1}{4}\xi_{xx}+\frac{1}{4}(7\xi_x+4\xi\frac{f_x}{f})_x=\frac{1}{4}\xi_{xx}$.
Substituting the last two expressions into the third equation of \eqref{ReguDEsystem6},
yields $r\,'(t)=0$ or $\xi=0$, which are the cases have already been discussed.\\

{\it Case 2.5:}  When $m\ne 0,1,2,3$, the third equation of \eqref{ReguDEsystem2} implies
\begin{equation}\label{ReguDEsystem7}
\begin{array}{ll}
\xi h\frac{f_x}{f}-\xi h_x +(1-m)h(2\xi_x+\xi\frac{f_x}{f})=0, \\
\xi_t=0, \\
(2\xi_x+\xi\frac{f_x}{f})_{xx}=0.
\end{array}
\end{equation}
Notice that the second equation of \eqref{ReguDEsystem7} indicates that $\xi$ is independent of $t$,
therefore the second equation of \eqref{ReguDEsystem2} is satisfied automatically.
Similar to the case of $m=0$, from the third equation of \eqref{ReguDEsystem7} and the first equation of
\eqref{ReguDEsystem2} we get the expression of $\xi$ and $f(x)$ as stated in  \eqref{ReguDEsystem3.4.1},
where $a$, $b$, $c$, $d$ are constants.
Substituting the expression of $\xi$ and $f$ into the first equation of \eqref{ReguDEsystem7} we obtain
\[
\displaystyle h(x)=\exp\Big(\int \dfrac{-7ax-7b+2c+(1-m)(ax+b)}{2ax^2+4bx-cx+d}\,{\rm d}\, x\Big).
\]
In addition, $\eta=2f\xi_t+(2\xi_x+\xi\frac{f_x}{f})u=(ax+b)u$. Therefore, we have
\begin{equation}\label{9RuguOperaOfEq}
\begin{array}{ll}
\xi=2ax^2+4bx-cx+d,  \\
\eta=(ax+b)u, \\
\displaystyle f(x)=\exp\Big(\int \dfrac{-7ax-7b+2c}
{2ax^2+4bx-cx+d}\,{\rm d}\, x\Big),\\
\displaystyle h(x)=\exp\Big(\int \dfrac{-7ax-7b+2c+(1-m)(ax+b)}{2ax^2+4bx-cx+d}\,{\rm d}\, x\Big).
\end{array}
\end{equation}
where $a, b, c, d$ are arbitrary constants. Thus, the corresponding regular reduction operator has the form
\[
Q=\partial_t + (2ax^2+4bx-cx+d)\partial_x+ (ax+b)u\partial_u,
\]
which is equivalent to Lie symmetry operator.

From the above discussion, we can arrival at the following two theorems.

\begin{theorem}\label{TheoremOfNontrivalRedOpe}
A complete list of $G_1^{\sim}$-inequivalent equations~\eqref{g11eqVarCoefNonLWaveEq} having nontrivial
regular reduction operator is exhausted by ones given in table~\ref{TableReRuOPClas}.
\end{theorem}

\setcounter{tbn}{0}

\begin{center}\footnotesize\renewcommand{\arraystretch}{1.15}
Table~\refstepcounter{table}\label{TableReRuOPClas}\thetable. Results of regular reduction operator classification of class \eqref{g11eqVarCoefNonLWaveEq} \\[1ex]
\begin{tabular}{|l|c|c|c|c|l|}
\hline
N & $n$ & $m$ & $f(x)$ & $h(x)$  & \hfil Regular reduction operator $Q$ \\
\hline
\refstepcounter{tbn}\label{CaseGrClasF1ForallHForAllK}\thetbn & $\neq-1$ & $n+1$ & $ \vert\, \xi \vert ^{-\frac{3n+4}{2n+2}}\exp \big(r\int \frac{1}{\xi}\,{\rm d} x\big)$ & $\dfrac{\xi_x^2-2\xi\xi_{xx}-p}{4(n+1)\xi^2}$ &
$\partial_t+\xi(x)\partial_x+(\frac{r}{n}+\frac{\xi_x}{2n+2})u\partial_u$ \\
\refstepcounter{tbn}\label{CaseF1HexpKexp}\thetbn & $1$ & $1$ & $1$ &
$b^2$ &
 $\partial_t + (\frac{1}{2}bx+\gamma_1e^{-bt}+\gamma_0)\partial_x+(bu-2\gamma_1be^{-bt})\partial_u$  \\
\refstepcounter{tbn}\label{CaseF1HexpK1}\thetbn & $1$ &$2$ &
$|\xi|^{-\frac{7}{4}}\exp(\alpha(t)\int\frac{{\rm d}x}{\xi})$ & $\frac{\xi_x^2-2\xi\xi_{xx}+q}{8\xi^2}$ & $\partial_t + \xi(t,x)\partial_x+\{2\xi_t|\xi|^{-\frac{7}{4}}\exp(\alpha(t)\int\frac{{\rm d}x}{\xi})$  \\
& $~$ & $~$ &
$~$ & $~$ & $+[\frac{1}{4}\xi_x+\alpha(t)]u\}\partial_u$  \\
\hline
\end{tabular}
\end{center}
{\footnotesize
Here $r, p, b, \gamma_1, \gamma_0$ are arbitrary constants, $\xi(x)$ in case 2.1 is an arbitrary functions of the variables $x$, $\xi(t,x)$ and $\alpha(t)$ in case 2.3
satisfy the overdetermined system of partial differential equations \eqref{7RuguOperaOfEq} and $\xi_t\neq 0$.
}

\begin{theorem}\label{TheoremOftrivalRedOpe}
Any reduction operator of an equations from class \eqref{g11eqVarCoefNonLWaveEq} having the form \eqref{GeneForReguOpe} with $\xi_t=0,~\xi_{xxx}=0$ is
equivalent to a Lie symmetry operator of this equation.
\end{theorem}

\section{Exact solutions}\label{SectiononLieandNonSymmRedu}
In this section, we construct nonclassical reduction and exact solutions for the classification models in table \ref{TableReRuOPClas} by using
the corresponding regular reduction operator. Lie reduction and exact solutions of equation from class \eqref{g11eqVarCoefNonLWaveEq} have been investigated
in reference \cite{Huang&Yang&Zhou2010a}. We choose case \ref{CaseGrClasF1ForallHForAllK} in table \ref{TableReRuOPClas} as an example to implement the reduction, the other cases can be considered in a similar way.

For the first case in table \ref{TableReRuOPClas}, the corresponding equation is
\begin{equation}\label{Cla1eqVarCoefNonLWaveEq}
[\vert\, \xi \vert ^{-\frac{3n+4}{2n+2}}\exp \big(r\int \frac{1}{\xi}\,{\rm d} x\big)]u_{tt}-(u^nu_x)_x-\dfrac{\xi_x^2-2\xi\xi_{xx}-p}{4(n+1)\xi^2}u^{n+1}=0,
\end{equation}
which admit the regular reduction operator
\[
Q=\partial_t+\xi(x)\partial_x+(\frac{r}{n}+\frac{\xi_x}{2n+2})u\partial_u.
\]
An ans\"{a}tze constructed by this operator has the form
\[
u(t,x)=\varphi(\omega)\vert\, \xi \vert ^{\frac{1}{2n+2}}\exp \big(\frac{r}{n}\int \frac{1}{\xi}\,{\rm d} x\big), \quad \mbox{where} \quad \omega=t-\int \frac{1}{\xi}\,{\rm d} x.
\]
Substituting this ans\"{a}tze into equation \eqref{Cla1eqVarCoefNonLWaveEq} leads to the reduced ODE
\begin{equation}\label{ExamRedODE1}
\begin{array}{ll}
[(4 r^2-p) n^2+4 (2 n+1) r^2] \varphi^{n+1}(\omega)
+4 n (n+1) [n \varphi''(\omega)-2 (n+1) r \varphi'(\omega)] \varphi^{n}(\omega)\\
+4 n^3 (n+1) \varphi'^{\;2}(\omega) \varphi^{n-1}(\omega)
- 4 n^2 (n+1) \varphi''(\omega)=0.
\end{array}
\end{equation}
Because there are higher nonlinear terms, we were not able to completely solve the above equation. Thus,
we try to solve this equation under different additional constraints imposed on $p$ and $r$.

We first rewrite equation \eqref{ExamRedODE1} as
\begin{equation}\label{ExamRedODE11}
\begin{array}{ll}
4 n^2 (n+1)[ \varphi'(\omega)\varphi^{n}(\omega)]'
-4 n^2 (n+1) \varphi''^(\omega)- 8 n (n+1)^2r \varphi'(\omega)\varphi^{n}(\omega)\\
+[4(n+1)^2 r^2-p n^2] \varphi^{n+1}(\omega)=0.
\end{array}
\end{equation}

If we take $p=4(1+\frac{1}{n})^2r^2$, then the general solution of \eqref{ExamRedODE11} can be written in the implicit form
\begin{equation}\label{Solu1ExamRedODE1}
\int \frac{n(\varphi^n-1)}{2r\varphi^{n+1}+c_1} {\rm d} \varphi=\omega+c_2.
\end{equation}
Up to similarity of solutions of equation \eqref{g11eqVarCoefNonLWaveEq}, the constant $c_2$ is inessential and can
be set to equal zero by a translation of $\omega$, which is always induced by a translation
of $t$.

If we further set $n=1$, the general solution \eqref{Solu1ExamRedODE1} can be  rewritten in the implicit form
\begin{equation}\label{Impli1Solu1ExamRedODE1}
\omega-\frac{\ln(2r\varphi^2-c_1)}{4r}-\frac{\sqrt{2}\arctanh(\frac{\sqrt{2}r\varphi}{\sqrt{c_1r}})}{2\sqrt{c_1r}}+c_2=0.
\end{equation}
Thus we obtain the following solution
\[
u(t,x)=\varphi(\omega)\vert\, \xi \vert ^{\frac{1}{4}}\exp \big(r \int \frac{1}{\xi}\,{\rm d} x\big), \quad  \quad \omega=t-\int \frac{1}{\xi}\,{\rm d} x
\]
for the equation
\[
[\vert\, \xi \vert ^{-\frac{7}{4}}\exp \big(r\int \frac{1}{\xi}\,{\rm d} x\big)]u_{tt}-(uu_x)_x-\dfrac{\xi_x^2-2\xi\xi_{xx}-8r^2}{8\xi^2}u^2=0,
\]
where $\varphi$ satisfy the equation \eqref{Impli1Solu1ExamRedODE1}, $\xi$ is an arbitrary function and $r$ is a non-zero constant.

If we further set $r = 0$, the general solution \eqref{Solu1ExamRedODE1} can be  rewritten in the implicit form
\begin{equation}\label{Impli2Solu1ExamRedODE1}
\frac{1}{n+1} \varphi^{n+1}(\omega)-\varphi(\omega)
=c_1\omega+c_2.
\end{equation}
Thus we obtain the following solution
\[
u(t,x)=\varphi(\omega)\vert\, \xi \vert ^{\frac{1}{2n+2}}, \quad  \quad \omega=t-\int \frac{1}{\xi}\,{\rm d} x
\]
for the equation
\[
\vert\, \xi \vert ^{-\frac{3n+4}{2n+2}}u_{tt}-(u^nu_x)_x-\dfrac{\xi_x^2-2\xi\xi_{xx}}{4(n+1)\xi^2}u^{n+1}=0,
\]
where $\varphi$ satisfy the equation \eqref{Impli2Solu1ExamRedODE1}, $\xi$ is an arbitrary function. In particular, for $n=1$ from equation \eqref{Impli2Solu1ExamRedODE1} we have
\[
\varphi(\omega)=1\pm\sqrt{1+2(c_1\omega+c_2)}.
\]
Thus we obtain an explicit solution
\[
u(t,x)=[1\pm\sqrt{1+2(c_1\omega+c_2)}]\vert\, \xi \vert ^{\frac{1}{4}}, \quad  \quad \omega=t-\int \frac{1}{\xi}\,{\rm d} x
\]
for the equation
\[
\vert\, \xi \vert ^{-\frac{7}{4}}u_{tt}-(uu_x)_x-\dfrac{\xi_x^2-2\xi\xi_{xx}}{8\xi^2}u^{2}=0.
\]
If we take different functions for $\xi$, then we can obtain a series of solutions for the corresponding equations. In order to avoid tediousness, we do not
make a further discussion here.

\section{Conclusion and Discussion}\label{SectionOnConclusion}
In this paper we have given a detailed investigation of the reduction operators of the variable coefficient nonlinear wave equations \eqref{eqVarCoefNonLWaveEq} (equivalently to \eqref{g11eqVarCoefNonLWaveEq}) by using the
singular reduction operator theory. A classification of regular reduction operators is performed
with respect to generalized extended equivalence groups. The main results on
classification for the equation \eqref{g11eqVarCoefNonLWaveEq} are collected in table~\ref{TableReRuOPClas} where we
list three inequivalent cases with the corresponding regular reduction operators. Nonclassical symmetry reduction of a class nonlinear
wave model \eqref{Cla1eqVarCoefNonLWaveEq} which are singled out the classification models are also performed.  This enabled to obtain some non-Lie exact
solutions which are invariant under certain conditional symmetries for the corresponding model.

The present paper is a preliminary nonclassical symmetry analysis of the class of
hyperbolic type nonlinear partial differential equations \eqref{eqVarCoefNonLWaveEq}.
Therefore, further investigations of different properties such as nonclassical potential symmetries,
and nonclassical potential exact solutions as well as physical application of this class of equations would be extremely interesting. These
results will be reported in subsequent publications.

\subsection*{Acknowledgements}
This work was partially supported by the National Key Basic Research Project
of China under Grant No.2010CB126600, the China Postdoctoral Science special Foundation under Grant No.201104247.

\end{document}